\documentclass[aps,pra,twocolumn]{revtex4}
\newcommand {\be}{\begin{equation}}
\newcommand {\ee}{\end{equation}}
\newcommand {\bey}{\begin{eqnarray}}
\newcommand {\eey}{\end{eqnarray}}

\usepackage{mathbbol}
\usepackage{epsfig}
\usepackage{amssymb}
\usepackage{amsfonts}
\usepackage{amsmath}
\usepackage{enumerate}
\usepackage{amsthm}
\usepackage{pgfplots}
\usepackage{tikz}
\usepackage{enumitem}

\usetikzlibrary{arrows.meta}

\usepackage{hyperref}

\theoremstyle{definition}

\newtheorem*{model}{The $\psi$-epistemic local model}

\newcommand{\mathsym}[1]{{}}
\newcommand{\unicode}[1]{{}}

\begin{document}

%\title{A local, many-worlds-inspired, model of quantum correlations with finite information flow}
\title{Local theories with parallel realities and the epistemic view of the quantum state}
\author{Alberto Montina, Stefan Wolf}
\affiliation{Facolt\`a di Informatica, 
Universit\`a della Svizzera italiana, 6900 Lugano, Switzerland}
\date{\today}

\begin{abstract}
Hidden-variable theories, such as the de Broglie-Bohm theory,
address the measurement problem by introducing auxiliary 
{\it random} variables that specify, in particular, the 
actual values of macroscopic observables. 
Such models may be $\psi$-epistemic, meaning the quantum state 
merely represents a state of knowledge, as opposed to $\psi$-ontic 
models, in which the quantum state is a physical variable of the theory.
A serious issue
of this route towards a realistic completion of quantum theory is raised
by Bell's proof that the resulting theories are nonlocal. A possible resolution
is to reject the assumption that measurements have single actual outcomes. 
Indeed, relaxing this premise, Deutsch and Hayden showed that 
Bell's theorem can be evaded
by delaying the buildup of the correlations until the parties compare 
their outcomes at a meeting point. However, the Deutsch-Hayden theory, which is 
deterministic and $\psi$-ontic, leads to an infinite information flow
towards the meeting point. Furthermore, alternative branches are weighted by amplitudes, 
leading to interpretative issues. 
In this paper, we introduce a general framework that combines the randomness of single-world 
theories with the coexistence of diverse instances, as found in many-worlds theory. This 
framework incorporates the existing theories as limiting cases.
We explore how this hybrid approach addresses key challenges 
of single-world and Deutsch-Hayden theories. On one hand, the multiplicity of coexisting 
instances allows us to circumvent nonlocality and, possibly, contextuality.
On the other hand, randomness makes it possible to derive 
quantum probabilities from unweighted counts of instances and ensemble averages.
Furthermore, it can lead to a reduction of the information flow by stripping
the quantum state of its `ontic' rank.
We illustrate this framework with a simple $\psi$-epistemic local model of projective 
measurements on two spatially separate maximally entangled qubits. Because of its randomness,
the model requires two unweighted instances and a finite information flow 
--  just one bit per measurement is communicated to the meeting point. 
Setting aside its foundational motivations, this framework has also relevance in quantum 
communication complexity and leads to novel technical questions, potentially providing 
new insights into some peculiarities of entanglement.
\end{abstract}
\maketitle
\section{Introduction}

The double-slit and Stern-Gerlach experiments are classic textbook 
examples that illustrate the core puzzle of quantum theory, that
is, the superposition and interference of distinguishable alternative states.
According to the Copenhagen interpretation, 
as long as a system is in a superposition of two or more states, 
none of them is actualized. Schr\"odinger famously illustrated the extreme 
implications of this view with his thought experiment, in which a cat is 
placed in a superposition of being both dead and alive~\cite{trimmer}. 
This scenario highlights a fundamental tension between the interpretation and our 
intuitive expectation that the cat must experience one definite state.

In modern terms, the apparent paradox in Schr\"odinger’s thought 
experiment finds a pragmatic resolution in decoherence theory. 
Since no experiment can detect interference between the "dead" 
and "alive" states, the superposition becomes {\it de facto} equivalent 
to a statistical mixture of the two. This allows for a consistent 
reduction to one of the two states without leading to contradictions. 
Thus, it is safe to claim that one of the alternatives is realized
with no future disproof of the claim.

While this approach leads to an interpretation of quantum
theory which is practically consistent, it also raises a
conceptual problem: how can one state in a superposition be actualized 
merely on the {\it promise} that no future observation will reveal the 
superposition? It would seem that the system has a kind of foresight of 
the future. For example,
the system could decohere by emitting a photon going into deep
space. Since it is plausible that the photon is lost forever, the
system could safely `decide' to reduce its state. However how does it
know that the photon will not encounter a mirror on one of Jupiter's
moons and be reflected back for re-absorption?

Since the inception of quantum theory, interpretative challenges have 
raised questions about its completeness~\cite{epr}. To reconcile state 
superpositions with the definiteness of our {\it unique} macroscopic
experience, a minimal requirement is the existence of auxiliary 
information that determines the actual macroscopic state that is 
experienced. This idea is pursued in ``hidden-variable''
theories and is consistently realized in the de Broglie-Bohm (dBB)
theory~\cite{bohm}. In the dBB theory, both the wave function and
the position of the particles specify the physical state of a system.
Employing a term coined in the context of quantum foundations, the
state represented by the wave function and the auxiliary variables is 
referred to as an {\it ontic state}, since it represents the
ontology of the theory. Generally, we refer to these hypothetical 
completions as {\it ontological theories}. In particular, theories in which 
the wave function is considered a physically real object are called $\psi$-ontic.
See Ref.~\cite{harrigan} for a comprehensive discussion on this classification.
It is important to emphasize that ontological theories are inherently 
{\it probabilistic}, reflecting the randomness and unpredictability of 
quantum phenomena. In Bohmian theory, for instance, particle positions 
are randomly distributed according to Born’s rule. 

However, this route towards a realistic completion of quantum theory leads
to a series of conceptual problems, such as contextuality~\cite{kochen} and, more 
seriously, nonlocality~\cite{bell}, on which we focus in this paper. As shown by 
Bell, the correlations between two entangled systems are incompatible with any
single-world
{\it local} ontological theory; a measurement performed on one system would
instantaneously affect the ontic state of the other, in apparent violation
of the principle of locality. 

The assumption of a single actual outcome of a measurement is an essential hypothesis 
of Bell's theorem. Indeed, Deutsch and Hayden (DH) developed a local many-worlds 
theory~\cite{deutsch} based on Everett's interpretation~\cite{everett}. In the latter, 
the wave function 
is identified as the complete ontology of quantum theory, so that its branching 
into alternative macroscopically distinct states is a physically real process. 
Thus, while single-world ontological theories are probabilistic, the many-worlds theory 
is $\psi$-ontic and deterministic. Randomness arises from the subjective perspective 
of an observer following a single branch of the tree.
By employing the Heisenberg picture, Deutsch and Hayden introduced additional 
ontological objects -- called {\it descriptors} -- in their local theory. Their 
argument for locality is clearly illustrated through a simple local realistic model of 
a  `nonlocal' Popescu-Rohrlich (PR) box~\cite{brassard}.
The core idea of the argument is to delay the buildup of correlations 
until the parties compare their outcomes at a meeting point. This delay
is possible only if multiple realities coexist locally.

However, since DH theory is $\psi$-ontic, there is an infinite flow of information 
from the parties to the meeting point. Furthermore, due to determinism, the theory 
faces the same issue as Everett’s: where do quantum probabilities come from? 
Although there is extensive literature on the derivation of Born’s rule, no 
broad consensus exists on how quantum probabilities emerge in Everett's interpretation.
Without entering 
into this debate, we simply assume -- within the context of the many-worlds 
theory -- that branches are weighted by an amplitude, and our expectation of 
being in a particular branch is given by Born’s rule, which we adopt as a 
postulate. Occasionally, we will also consider the branch counting proposed
by Saunders~\cite{saunders}, since this has some similarity with the counting
that we will employ in the framework discussed in this work. 
Motivated by both foundational considerations and communication-complexity questions, 
our aim is to investigate alternative local theories that employ fewer resources and 
unweighted counting rules. To this end, we introduce a general framework that 
encompasses both single-world and many-worlds theories as limiting cases.

In the search for an alternative local theory with multiple instances,
it is fundamental to emphasize that the branching of the wave function constitutes a 
physical process in both the dBB and Everett theories. In the dBB theory, however, the 
positions of the particles determine which branch is actually experienced. For this 
reason, some proponents of the many-worlds interpretation argue that dBB is effectively 
a many-worlds theory augmented with additional (and, in their view, unnecessary)
variables. See, for example, the discussion in Ref.~\cite{brown} and references therein. Indeed, 
it may seem peculiar that alternative branches of the wave function -- 
each containing detailed biophysical processes, such as brain activity -- exist physically, 
yet are empty of observers. These brain activities in ``empty'' branches would be
akin to ``philosophical zombies'' introduced in philosophy of mind and in debates about
consciousness. 

With this important premise on the dBB theory, one might wonder whether any completion 
of quantum theory is essentially just a many-worlds theory augmented with extra-variables.
The crucial point of the question lies in the 
interpretation of the wave function. If the wave function is regarded as part 
of the ontology, then it makes sense to claim that the various branches 
of the wave function correspond to physically existing realities, even if 
observers, attached to the extra-variables, may follow only one of them.
Conversely, each branch of the wave function does not necessarily have physical
reality in specific ontological theories, known as $\psi$-epistemic 
theories, which have gained significant interest over the past 
decade~\cite{pbr,hardy,leifer0,barrett,leifer,ring}. In this 
subclass,  the quantum state is interpreted as a state of knowledge 
and is not part of the ontology; instead, its information 
is encoded in the statistics of the ontic state. Because of this
encoding, $\psi$-epistemic theories are essentially equivalent to
classical simulations of quantum communication which employ finite
classical communication~\cite{montina_comm_epist0,montina_comm_epist}.
In a $\psi$-epistemic ontological theory, the coexistence of multiple realities 
would not be justified by the branching of the wave function, but rather
by Bell's theorem. Furthermore, in such theories, the branching of the wave 
function would not necessarily imply a branching into distinct realities.
Since $\psi$-epistemic ontological theories are the only alternative to 
$\psi$-ontic many-worlds theories with possible extra-variables, we focus
on them. An exception worth noting is the {\it many interacting worlds} (MIW)
approach of Ref.~\cite{hall}, which is $\psi$-ontic in some sense, as discussed
later. However, it is not equivalent to many-worlds theory, since the equivalence
to quantum mechanics holds only in the limit of infinite worlds (an interesting
question regarding this approach will be discussed in the conclusions).

In this paper, we explore whether integrating the randomness inherent in single-world 
ontological theories with the coexistence of multiple instances -- as in the many-worlds 
theory -- can yield a more general framework, while preserving the locality of the 
Deutsch–Hayden (DH) theory. In particular, the introduction of randomness is a 
prerequisite for a finite information flow, which is infinite in the DH theory. 
Moreover, relaxing the determinism of the many-worlds theory makes it
possible to derive quantum probabilities through a simple unweighted count of instances 
and an ensemble average.
We illustrate this hybrid framework that incorporates both randomness 
and multiplicity of instances
by showing that just one classical bit of information about the chosen
 measurement and two unweighted instances are sufficient to simulate 
local projective measurements on maximally entangled qubits.
This model raises the question whether a local $\psi$-epistemic
theory can be built. 

Our model directly builds on the Toner-Bacon model, which simulates 
maximally entangled states using one bit of communication and shared 
randomness~\cite{toner}. By selecting an instance at random, our model 
reproduces the same statistics as the single-world model in Ref.~\cite{gisin}, 
in which outcomes are generated through shared randomness and the single
use of a nonlocal PR box.

In Sec.~\ref{onto_model_sec}, we discuss the general framework of a single-world
ontological theory. In Sec.~\ref{many_real_sec}, the framework is
extended to the case of multiple coexisting realities. This extension includes 
many-worlds theory as limiting case. In Sec.~\ref{localMWI_sec},
a simplified version of DH theory is discussed in the two-party Bell scenario.
In Sec.~\ref{our_model_sec}, we introduce our model with finite information flow.
A discussion follows in Sec.~\ref{discussion_sec}. Finally, we draw the conclusions
and perspectives. In particular, we remark that the introduced framework offers
a generalized setting for quantum communication complexity, which has its own
appeal beyond foundational motivations. 

\section{Single-world ontological theories}
\label{onto_model_sec}
Ontological theories of quantum mechanics have been motivated by the apparent
contradiction between the well-defined macroscopic reality and 
the linear evolution of the wave function, leading to the measurement problem.
Since the cat in Schr\"odinger's thought experiment must be dead or alive
after all, some auxiliary information about his actual state should be added to
the information contained in the wave function, provided that the linearity of the
Schr\"odinger equation is assumed universal. This completion is consistently
realized, for example, in the dBB theory~\cite{bohm}, in which both the
wave function and the position of the particles are part of the ontology.
In this respect, the dBB theory could be considered a kind of many-worlds 
theory~\cite{brown} with extra-variables labeling which branch is experienced.
Is any completion of quantum theory just a many-worlds theory augmented with 
extra-variables? That is, should the wave function be considered a physically
real entity in any ontological theory?
In the context of quantum foundations, such completions, typically referred to 
as $\psi$-ontic, are opposed to $\psi$-epistemic theories, where the wave function represents 
only an observer’s knowledge or information about a system. This distinction is
extensively discussed in the seminal paper by N. Harrigan and R. W. Spekkens~\cite{harrigan},
who advocate for the $\psi$-epistemic view. Let us quote the closing lines of their abstract: 
``[...] Einstein was seeking not 
just any completion of quantum theory, but one wherein quantum states are 
solely representative of our knowledge''. In particular,
the authors point out that one of Einstein's arguments for incompleteness of
quantum mechanics leads to support the epistemic view of quantum states under the 
premise of local causality. Unfortunately, Einstein's argument alone is not supportive 
of the $\psi$-epistemic view, since local causality contradicts quantum mechanics 
under the condition of single actual outcomes,
as shown by Bell. However, there are other reasons for promoting a $\psi$-epistemic
view~\cite{spekkens}. Furthermore, $\psi$-epistemic models have an important role
in quantum communication complexity, since they are essentially equivalent to 
classical simulations of quantum communication that employ finite classical 
communication~\cite{montina_comm_epist0,montina_comm_epist}. Since the 
$\psi$-epistemic view is the only possible alternative to an ontological theory 
which is not just many-worlds theory with extra-variables, it merits further 
consideration.

An ontological theory of quantum phenomena is characterized by an ontological space $\Omega$ 
of ontic states, denoted by $\lambda$, which are interpreted as physically real states 
of the system. Generally, ontological models are introduced in the context of a quantum 
system $S$,
which undergoes a preparation and a subsequent measurement~\cite{harrigan}. In that context, 
the ontic state refers to the quantum system $S$, whereas measurements are described by indicator
functions, which are conditional probabilities of getting an outcome given $\lambda$. 
As our focus is on a theory rather than a model, we include macroscopic systems interacting 
with $S$ in the description.
Thus, the ontic state contains, in particular, full information about macroscopic facts,
such as the position of the pointer of a measuring device.
For example, in Schr\"odinger's thought experiment, there exists a subset $\Omega_A\subset\Omega$ of 
the ontological space such that $\lambda\in\Omega_A$  if and only if the cat is alive.

Ontic states are typically assumed to evolve deterministically, or more generally, according 
to some causal rule, such as a Markov process. Since quantum theory is fundamentally
probabilistic, the preparation of a quantum 
state $\psi$ does not uniquely determine the ontic state $\lambda$, but rather specifies a probability 
distribution $\rho(\lambda|\psi)$, possibly conditioned on additional contextual parameters, that
is, we have the mapping
\be
\psi \rightarrow \rho(\lambda|\psi).
\ee
Crucially, the converse is also true in $\psi$-epistemic ontological theories, that is,
the ontic state $\lambda$ does not uniquely determine the wave function. This
is equivalent to state that the wave function is not an element of the ontology.
Within this framework, the branching of the wave function is not necessarily a physical process, 
since the wave function itself is not taken to represent a physical entity. Instead, its role 
is epistemic: its information is statistically encoded in the distribution of the underlying ontic state. 
Suggesting that a wave function branches into multiple realities would, in this view, be akin to 
claiming that a tossed coin splits into two worlds merely because each outcome is possible.
Thus, a $\psi$-epistemic ontological theory is not a many-worlds theory with extra-variables.
The coexistence of multiple realities will be an additional hypothesis justified by Bell 
nonlocality, rather than the branching of the wave function.

\section{Ontological theories with multiple coexisting realities}
\label{many_real_sec}

Curiously, the term $\psi$-epistemic has traditionally been used only in the context of single-world 
theories. In contrast, within the many-worlds community, the view that the wave function has 
ontological significance appears to be ubiquitous. For example, in Ref.~\cite{kent}, the ontological 
status of $\psi$ is identified as one of the two essential characteristics of the galaxy of 
many-worlds interpretations. Kent states: `` DeWitt (writes) `[...] The symbols of quantum mechanics 
represent reality just as much as do those of classical mechanics.' There seems to be no dispute in 
the literature on this point: if a theory is not mathematically realist then it is not a many-worlds 
interpretation.'' Indeed, the ontology of the wave function motivates the branching into
many parallel worlds. 
As far as we know, the epistemic view has never been considered in a scenario involving multiple 
coexisting realities. Remarkably, the model presented in Ref.~\cite{hall} replaces the wave function 
with many interacting worlds, so that the quantum state is not treated as an element of reality 
in the theory. However, the model -- which is nonlocal -- reproduces quantum mechanics exactly only in 
the limit of infinitely many worlds. In this limit, the collective classical state of the infinite worlds 
contains complete information about the quantum state. Thus, the theory becomes $\psi$-ontic as the 
discrepancy with quantum predictions vanishes. 

In this section, we extend the framework of single-world ontological
theories by assuming that multiple realities  may coexist. These theories are necessarily
probabilistic as long as the number of realities is kept finite.
This contrasts with the many-worlds theory, where probabilities must be justified or postulated 
a posteriori. The determinism of many-worlds theory can be recovered in the limit of infinite realities. 
As previously said, if the considered framework is 
$\psi$-epistemic, then the branching is not motivated by the branching of the wave function, rather by
Bell's theorem. Our aim is to explore whether quantum correlations can be understood in a locally 
causal framework, simply by relaxing the assumption of a single realized world without adopting
the DH many-worlds theory as a whole. 

This section is organized as follows. In Sec.~\ref{multiple_instances_sec}, we extend the
single-world ontological theories by introducing multiple coexisting instances of
some macroscopic physical quantity. In Sec.~\ref{instances_branches_sec}, we clarify
the terms {\it instances} and {\it branches} in our context by introducing an alternative 
representation using occupation numbers, which is more reminiscent of  the formalism
of many-worlds theory, in particular of Saunders' branch counting. Furthermore,
we show that many-worlds theory can be recovered (with some extra structure) 
in the limit of infinite instances.
In Sec.~\ref{2macro_quant_sec},
the combination of two macroscopic quantities is discussed, which is preparatory for
the definition of {\it separability} and {\it locality} of the theory, given in 
Sec.~\ref{sec_locality}.

\subsection{Multiple instances of a macroscopic quantity}
\label{multiple_instances_sec}
Let us consider a laboratory with a quantum system ${\cal Q}$, a measuring device and an observer,
say Alice. The device has a pointer moving to some position after a von Neumann measurement on 
$\cal Q$ is performed. This position is observed by Alice. The position of the pointer (as well as
the associated neural activity of the observer) is referred to as a macroscopic physical quantity
$\cal M$.
Under the assumption of a single actual world, the ontic state determines one single value $s$ of 
${\cal M}$, so that the variable $s$ is part of the ontology 
of the theory. Relaxing this assumption, we assume that there are $m$ variables
$s_1,\dots,s_m$ representing the values of $\cal M$ in each of $m$ realities or {\it instances}.
The ultimate goal is to find an $m$ such that the theory is local. Bell's theorem implies that 
$m>1$. Additionally, we include some other variables, which we denote by $\eta$.
The overall set of instances is described by the ontic state 
\be\label{ontic_instances}
\lambda=(s_1,s_2,\dots,s_m,\eta),
\ee
which is an element of the ontological space $\Omega$.
More generally, the physical quantity $\cal M$ is associated
with a partition of $\Omega$ into subsets $\Omega_{s_1,s_2,\dots,s_m}$ such that 
$\lambda\in \Omega_{s_1,\dots,s_m}$ if and only if $\cal M$ is equal to $s_k$ in the
$k$-th instance for every $k\in\{1,\dots,m\}$.

Since the preparation 
of the quantum state does not uniquely determine the ontic state $\lambda$, the $m$-tuple of outcomes 
$(s_1,s_2,\dots,s_m)$ generally emerges from the measurement with probability 
$\rho(s_1,s_2,\dots,s_m)$. Like 
in many-worlds theory, a single observer will experience one of the $m$ outcomes.
In a single-world theory, one would have a distribution $\rho(s)$ over a single outcome $s$, 
consistent with Born’s rule. In our $m$-instances scenario, however, we must introduce an 
additional assumption to connect the joint distribution $\rho(s_1,s_2,\dots,s_m)$ with the empirical data 
observed by a single observer:
\newline
{\bf Assumption} {\it (Equiprobability of instances)}.
The observer is equally likely to be in either reality, 
without any weight attached to the instances.

This is motivated by the principle of 
indifference (also known as the principle of insufficient reason). The same assumption
has been made in Refs.~\cite{hall,brassard}.
Accordingly, the probability of observing outcome $s$ is 
\be
\rho(s)=\frac{1}{m} \sum_k \sum_{s_1,\dots,s_m} \delta_{s_k,s}\rho(s_1,\dots,s_m)
\ee
This is equivalent to a simple unweighted counting rule: we count how many of the $m$
realities yield outcome $s$ (from $0$ to $m$ realities), average over many runs, 
and divide by $m$. This evaluated $\rho(s)$ must agree with Born's rule.

Two points are crucial here. First, the number of instances with given outcome $s$
fluctuates from run to run, such that the observed frequencies converge to those predicted 
by quantum theory as the number of runs goes to infinity. If the theory were deterministic, one
would observe only frequencies equal to a rational number in $\{0,1/m,2/m,\dots,1\}$ 
(the latter with $s=s_1=s_2=s_3=\dots=s_m$). Second, in each individual run, not 
all outcomes need to be realized, in contrast to the many-worlds framework, where all the branches
of the wave function
are instantiated. In our example, at most $m$ distinct outcomes can occur in a single run (only one if 
$s_1=s_2=\dots=s_m$). If $m < n$, where $n$ is the number of branches of the wave function, we would 
have at most $m$ distinct outcomes per run, not the $n$ outcomes 
realized in the many-worlds theory. Even if $m>n$, the number of distinct realities may be smaller
than $n$ in some runs of the experiment. In particular, if there exists a branch of the wave function 
with an extremely small probability $p$, 
this means that, based on our knowledge of the system, there is an extremely small probability 
(not greater than $m p$) that some
instance realizes that branch. Thus, it is not guaranteed that such an instance exists in every run of 
the experiment. This is trivially true in a single-world ontological theory with only one existing 
reality. This departs sharply from the deterministic realization of all the branches in many-worlds 
theories. Provided that the ontic state $\lambda$ does not contain $\psi$, the framework
also departs from dBB theory, where there are branches which are physically real, but empty
of observers.

It is trivial to prove that this framework is consistent with quantum mechanics.
In the de Broglie-Bohm theory with one instance ($m=1$), $\eta$ is identified with the wave
function and $s_1$ represents the position of all the particles in the laboratory (including
the particles of the pointer). A naive many-instances theory can 
be trivially built by introducing $m$ trajectories in the pilot-wave theory such
that the position of the particles in each instance is distributed according to Born's rule. 
Whereas such many-instances extensions of the de Broglie-Bohm theory shows that 
the framework does not conflict with quantum mechanics, they have no appealing
motivation. The dBB theory consistently solves the measurement problem (which was
the motivation of the theory), but raises new problems (nonlocality~\cite{bell}) and critics (existence
of ``empty'' branches~\cite{brown}). Introducing additional trajectories does not weaken these
new concerns. It is decisively more interesting to investigate $\psi$-epistemic
theories of multiple instances, in which $\eta$ is not identified with the quantum
state or, more generally, $\lambda$ does not completely define the quantum
state.

At this point, it may be useful to clarify the terms {\it instances, realities and branches} 
used in our context. This will further highlight the similarities between our general 
framework and many-worlds theory -- the latter being a limiting case of the former. 
Later, we will discuss how to combine
different macroscopic quantities. This will allow us to introduce the concepts of
{\it separability} and {\it locality}.

\subsection{Branches versus instances}
\label{instances_branches_sec}
Branches, worlds, and realities are interchangeable terms 
in the context of Everett's many-worlds theory. When the wave function goes into 
a superposition of macroscopically distinct states, we say that one world branches
into separate worlds or macroscopically distinct realities. Thus, by definition,
two branches, worlds or realities differ by some physical quantity. For example, if 
the physical quantity has $n$ different values and each value has a nonzero probability 
amplitude, we say that we have $n$ different branches or worlds, although they may have 
different weights.

In our context, there is a distinction between realities, or {\it instances}, and the 
branches identified by the values taken by physical quantities. Instances are
meant as realities with their own private ontic variables $s_k$,
but these variables may take identical values in different instances.
Consequently, we say that the system branches into two distinguishable realities when 
some physical quantity in different instances evolves from one shared initial value 
to two distinct values. Let us further clarify this
point by discussing an alternative representation of the ontological theory,
which aligns more closely with the narrative of the many-worlds theory.

In the description of Eq.~(\ref{ontic_instances}), instances are labeled by some index, 
akin to the particle description in first quantization, in which the vector position 
$\vec x_k$ of a particle has an underscript $k$ identifying the particle. 
However, a more concise description of a set of identical particles is provided
in second quantization, that is, by the quantization of a field. Rather than saying
that particles $k_1, k_2,\dots, k_h$ are in the state $s$, we just say that
the state $s$ has occupation number $h$. Similarly, if the description of $m$ instances 
is invariant under a permutation of the labels, we can just specify the number of instances 
in which a physical quantity $\cal M$ takes any value $s$. Let $\{1,\dots,M\}\equiv \Gamma$
be the set of possible values of $\cal M$, the ontic state in Eq.~(\ref{ontic_instances})
would be replaced by
\be\label{ontic_with_occupation}
\lambda=(n_1,\dots,n_M,\eta),
\ee
where $n_i$ is the number of instances (occupation number) with ${\cal M}=i$. Also in
this case, the description can be generalized by partitioning the ontological space
into subsets $\Omega_{n_1,n_2,\dots,n_M}$ such that $\lambda\in\Omega_{n_1,n_2,\dots,n_M}$
if and only if $n_i$ instances have ${\cal M}$ equal to $i$ for every $i\in\{1,\dots,M\}$.

The new representation using occupation numbers is more akin to the usual picture
of a many-worlds theory. The label $i$ identifies a branch, whereas the occupation
numbers, which are integers, replace the quantum probabilities weighting the
branches of many-worlds theory. 
Provided that the number of branches exceeds the total occupation number,
it is clear that many branches have zero occupation number (zero instances).
In a single-world ontological theory, the total occupation number is just one,
so that only one branch is realized.

Our weighting of branches with an occupation number also somehow resembles 
Saunders' branch counting~\cite{saunders}. Saunders argues that the quantum
states continuously undergo a huge number of branching at the microscopic
level. These microscopically distinguishable states result into few 
indistinguishable macroscopic states. 
By partitioning macroscopic branches into suitable sets 
of microscopic branches with equal weight, Born’s rule is recovered 
through a simple count of microscopic branches in the limit of infinite partitioning.
Saunders' microstates and macrostates are akin
to {\it instances} and {\it branches} in our terminology. As in our case, each value 
of a macroscopic observable is weighted by an integer. The mark difference is that 
Saunders' weights must go to infinity to recover Born's rule because of determinism.
Conversely, the weights $n_i$, which are stochastic variables, 
are not proportional to the quantum probabilities in a single run -- Born's rule is 
recovered only on average. This is what happens in every single-world ontological
theory, in which only one value has occupation number equal to $1$, whereas
the others are equal to zero. The value taken by the single instance 
changes stochastically from run to run. Consider, for example, the de Broglie-Bohm 
theory, in which the particle positions are probabilistically distributed according 
to Born's rule. 

\subsubsection{Many-worlds theory in the limit $m\rightarrow\infty$}
\label{limit_sec}
The set of theories fitting into the general framework previously introduced 
is not empty and includes every single-world ontological theory with
$m=1$. These theories, such as dBB theory,
can be trivially extended to theories with multiple realities with $m>1$.
Furthermore, the framework also includes
many-worlds theory in the limiting case of $m$ going to infinity (with some extra structure), 
as discussed in the following. First of all, we take for granted that there is a preferred basis in
many-worlds theory on which the branching occurs and this basis is defined
by the eigenvectors of $\cal M$. 
The first step is to identify $\eta$ in Eq.~(\ref{ontic_with_occupation})
with the quantum state. This makes the framework $\psi$-ontic. The second step is
to replace the occupation numbers $n_i$ with $\bar n_i\equiv n_i/N_t$, where $N_t$ is the
total number of instances. The rational number $\bar n_i$ is trivially the probability of
being in branch $i$ in a single run. This probability is bounded for every $N_t$, so
that we can take the limit $N_t\rightarrow\infty$. At this point, we impose determinism
and set
\be\label{constr_occup}
\bar n_i=|\psi_i|^2
\ee
in each run. The quantity $\bar n_i$ is the fraction of instances in the branch $i$, which
resembles Saunders' branch counting. This quantity offers a convenient 
way to explain why branches
with greater weights are more likely to be experienced -- they are associated with
a higher multitude of instances, so that the previously stated assumption of {\it equiprobability
of instances} holds. 
In the limit $N_t\rightarrow\infty$, the framework
with constraint~(\ref{constr_occup})
is {\it de facto} equivalent to many-worlds theory with the quantities $\bar n_i$ as
extra structure. The constraint~(\ref{constr_occup}) may be justified by the 
diverse derivations of Born's rule given in many-worlds theory. For example,
the constraint can be derived by assuming that $\bar n_i$ is a function of
$|\psi_i|$ and that an additivity property holds, along the same lines as 
Everett~\cite{everett}. Namely, the additivity property states that
the probability weight of $\sum_k a_k |\phi_k\rangle$ is equal to the sum of 
the probability weights of $a_1,a_2,\dots$, where $|\phi_1\rangle,|\phi_2\rangle,\dots$ are
orthonormal vectors. This property guarantees consistency of the constraint
under unitary evolutions -- namely, it ensures the normalization of the
probabilities over time.

It can be interesting to investigate whether constraint~(\ref{constr_occup}) can be imposed 
only on average by retaining the stochasticity of the framework in the limit 
$N_t\rightarrow\infty$.

\subsubsection{Conservation of the number of instances}
The representation using occupation numbers might raise the question of whether the 
total number of instances can be created or annihilated, analogous to particles in 
quantum field theory. Given the analogy between occupation numbers and the quantum 
probabilities weighting the branches, the conservation of the total number of instances 
appears as a natural hypothesis, mirroring the conservation of the norm of the wave 
function. Indeed, the creation of new instances would correspond to the ex novo 
creation of macroscopic systems -- such as observers and measuring devices. This would 
be peculiar, especially considering that each instance is regarded as a physically 
existing system. Such creation would imply an increase in energy. Indeed,
branching in the many-worlds theory does not involve the creation of new systems, 
but rather a divergence into different states of the same system. Similarly,
in our framework, branching corresponds to a fixed number of instances whose states
diverge from a shared initial state.

\subsection{Combination of two macroscopic quantities}
\label{2macro_quant_sec}
Before introducing {\it separability} and {\it locality} in the theory, let us discuss
the rule for combining two macroscopic quantities, say ${\cal M}_A$ and ${\cal M}_B$, generated
by performing two different measurements on the quantum system ${\cal Q}$. Each of these 
quantities are, for example, the pointer of some measuring device. There are
different ways of generating ${\cal M}_A$ and ${\cal M}_B$. Let us analyze three cases:
\begin{enumerate}
\item One observer, say Alice, performs first one measurement generating the macroscopic quantity ${\cal M}_A$,
then she performs a second measurement generating ${\cal M}_B$. 
\item 
Alice performs a measurement, generating ${\cal M}_A$, then departs from the measured system. 
Subsequently, another party, Bob, performs a second measurement, generating  ${\cal M}_B$.
Alice and Bob have not interacted with each other in the interval between the measurements.
\item Given two entangled systems ${\cal Q}_A$ and ${\cal Q}_B$ that are spacelike-separated, Alice and Bob 
perform a local measurement on ${\cal Q}_A$ and ${\cal Q}_B$, respectively. Eventually, they
meet to compare their results.
\end{enumerate}

Let us assume that there are $m$ realities of Alice.
In the first case, when Alice performs the first measurement, the pointer of one measuring
device in the $k$-th instance moves to the position $s_{A,k}$, which is observed 
by Alice. After a second measurement is performed, the pointer of another device moves to 
position $s_{B,k}$, which is again observed. Since the number of
realities of Alice is preserved, in the end, the overall quantity 
${\cal M}={\cal M}_A \oplus {\cal M}_B$ has $m$ instances with value $(s_{A,k},s_{B,k})$ 
in the $k$-th instance. Thus, the value $s_k$ in Eq.~(\ref{ontic_instances})
is identified with the pair $(s_{A,k},s_{B,k})$. It is worth to note that the combination
of the quantities corresponds to a direct sum, rather than a tensor product.

The second case is more intricate. Let us assume that Bob has a number $m'$ of instances.
Since Alice and Bob never interact in the interval between the measurements,
all we can say immediately after the last measurement is that Alice has $m$ instances of 
$s_A$ and Bob $m'$ instances of $s_B$. There is no reason for pairing their respective 
outcomes as long as they stay far away. The ontic state is
\be
\lambda=(s_{A,1},s_{A,2},\dots,s_{A,m},s_{B,1},s_{B,2},\dots,s_{B,m'},\eta),
\ee
in which the ordering of $s_A$ and $s_B$ establishes no pairing rule of Alice and Bob's instances.
In this respect, Peres' dictum "unperformed measurements have no result"~\cite{peres} holds. 
Only when Alice and Bob meet, a pairing of their instances is defined. Since the number of
instances of each party is preserved and one party eventually experiences 
one definite macroscopic state of the other party, the pairing must be one-to-one, so 
that $m'=m$. For example,
if Alice had fewer instances, some of them would need to pair
with more than one of Bob’s. Such a mismatch would generally require the creation of
Alice’s instances. 

The specific way in which the pairing occurs depends on the ontic state at the moment 
the meeting takes place. The pairing is understood as a physical process in which one 
instance of Alice interacts with only one instance of Bob. Note that if the theory is 
deterministic, the ontic state contains information about the pairing even prior to 
the meeting of the parties. It is akin to saying that a neutron and a proton will bind 
if they are brought sufficiently close together. We seek a local mechanism for the 
pairing, which, however, may be mathematically defined in a nonlocal way at any given time.

The framework may be generalized to infinite instances. In this generalization, it is required 
that the cardinality of the set of instances is identical for each quantity (e.g. countable, 
continuum and so on).

Summarizing, we have the following properties
\begin{itemize}
\item Each macroscopic quantity has a fixed number $m$ of instances (more generally, the
set of instances has fixed cardinality).
\item If two quantities ${\cal Q}_A$ and ${\cal Q}_B$ are jointly known, then
there is a one-to-one correspondence between the instances of ${\cal Q}_A$
and the instances of ${\cal Q}_B$. This correspondence defines a combination
${\cal Q}_A\oplus {\cal Q}_B$ with $m$ instances. The correspondence is determined by the ontic 
state $\lambda$.
\end{itemize}

The third case can be treated in the same way and is discussed in the following subsection.

\subsection{Separability and locality}
\label{sec_locality}
Until now, we have introduced a general framework of an ontological theory with
multiple realities. As previously said, the generalization to multiple instances
is suggested by Bell's theorem, which is proved under the assumption of single
actual outcomes. Thus, we are particularly interested to formulate a {\it separable}
and {\it local} framework. Here, we will consider the Bell scenario of two 
spacelike-separated laboratories (Case 3 of Sec.~\ref{2macro_quant_sec}).
Each laboratory includes a system (${\cal Q}_A$ and ${\cal Q}_B$), a measuring device, 
and an observer -- Alice and Bob. The device has a pointer that is set in a ready state 
and indicates an outcome $s$ after a measurement is performed on the system. The two 
systems ${\cal Q}_A$ and ${\cal Q}_B$ are generally in some entangled state. Eventually, Alice
and Bob compare the results at a meeting point. Before the measurements, the instances
are macroscopically indistinguishable, so that there is only
one macroscopically distinct realization of Alice and Bob. As previously said,
to avoid the creation of new instances over time, we assume that each spatially 
distinct object has the same number $m$ of instances. 

{\bf Separability} --
Like in a separable one-world ontological theory, the two laboratories have their 
own ontic states, say 
\be
\label{local_ontic}
\left\{
\begin{array}{l}
\lambda_A=(s_{A,1},s_{A,2},\dots,s_{A,m},\eta_A), 
\vspace{1mm}  \\
\lambda_B=(s_{B,1},s_{B,2},\dots,s_{B,m},\eta_B).
\end{array}
\right.
\ee
It is important to emphasize again that the $m\times m$ possible combinations of 
local instances do not correspond to $m\times m$ instances of the overall system composed 
of the two laboratories. Each party has exactly $m$
instances at all times -- including when they meet to compare their results. At the meeting point, 
one instance of Alice will interact with exactly one instance of Bob, and vice versa. Thus, given 
the one-to-one pairing rule, there will be $m$ pairs of Alice and Bob after they interact.

{\bf Locality} -- Initially, at time $t=0$, the ontic states are 
$\lambda_A'=(s_{A,1}',s_{A,2}',\dots,s_{A,m}',\eta_A')$ and 
$\lambda_B'=(s_{B,1}',s_{B,2}',\dots,s_{B,m}',\eta_B')$,
$s_{A,k}'$ and $s_{B,k}'$ being in the ready state $0$. Later, Alice and Bob perform 
each a measurement randomly selected among a set of possibilities. Let us denote
by parameters $a$ and $b$ their respective selection. As usual in Bell's scenario,
we assume that $a$ and $b$ are not correlated with $\lambda_A'$ and $\lambda_B'$. These
parameters are randomly generated outside the laboratories and passed to Alice and
Bob.  The ontic states evolve to 
$\lambda_A''=(s_{A,1}'',s_{A,2}'',\dots,s_{A,m}'',\eta_A'')$ and 
$\lambda_B''=(s_{B,1}'',s_{B,2}'',\dots,s_{B,m}'',\eta_B'')$,
where  $s_{A,k}''$ and $s_{B,k}''$  take the value of the outcome in
the $k$-th instance of Alice and Bob, respectively.

At this point, we assume that the theory is {\it locally causal}~\cite{bell76}. On Alice side we 
impose that the probability of getting $\lambda_A''$  given $a$, $b$, $\lambda_A'$, $\lambda_B'$
and $\lambda_B''$ does not depend on $b$, $\lambda_B'$ and $\lambda_B''$. Similarly, we impose a specular 
condition on Bob side. That is,
\be\begin{array}{l}
\label{local_causal_eqs}
\rho(\lambda_A''|\lambda_B'',\lambda_A',\lambda_B',a,b)=
\rho(\lambda_A''|\lambda_A',a)  \vspace{1mm} \\
\rho(\lambda_B''|\lambda_A'',\lambda_A',\lambda_B',a,b)=
\rho(\lambda_B''|\lambda_B',b).
\end{array}
\ee
These conditions lead to the factorization 
\be\label{factor_eq}
\rho(\lambda_A'',\lambda_B''|\lambda_A',\lambda_B',a,b)=
\rho(\lambda_A''|\lambda_A',a)  
\rho(\lambda_B''|\lambda_B',b),
\ee
from which Bell's inequalities are derived in the single-world case~\cite{bell76}. Generally,
$\lambda_A'$ and $\lambda_B'$ are correlated, as the systems are entangled. Under the 
hypothesis of determinism, the assumption of local causality can be replaced by the
weaker assumption of {\it locality}. See Ref.~\cite{wiseman} for a comprehensive discussion
on the conceptual differences between the two assumptions in the context of Bell's proofs.

{\it Comparing the outcomes} -- After the measurements, Alice and Bob reach the meeting 
point and compare their results.
The $i$-th instance of Alice is paired with the $k_i$-th instance of Bob, where
$k_i=k_i(\lambda_A'',\lambda_B'')$ is some bijective function in $i$ depending on
the ontic states $\lambda_A''$ and $\lambda_B''$. Depending on the model, the
pairing may need only a partial information on $a$ and $b$, which is encoded
in $\lambda_A''$ and $\lambda_B''$. Conversely, the DH theory
needs the full information on the setting parameters (as shown later in Sec.~\ref{localMWI_sec}).
In Sec.~\ref{our_model_sec}, we will see that the correlation of two maximally
entangled qubits can be reproduced by a local model in which the 
two parties need to carry just one bit of information about their performed
measurement to the meeting point. Furthermore, the model requires just two
instances per party. In general,
the number of realities and the amount of information can depend on the dimension of the 
Hilbert space of each subsystem, the number of subsystems and the computational network 
within which the subsystems interact. 

In the next section, we will introduce a simplified version of the DH theory applied to
the two-party Bell scenario, then we will introduce the $\psi$-epistemic model.
\section{Nonlocal correlations in many-worlds theory}
\label{localMWI_sec}
In Ref.~\cite{deutsch}, Deutsch and Hayden proposed a local, many-worlds interpretation of 
quantum correlations by introducing a set of operators -- referred to as 
{\it descriptors} -- 
that evolve in the Heisenberg picture. This framework significantly expands the ontology of 
the theory by incorporating these new entities.
Here, we present a highly simplified caricature of their argument. While this oversimplification 
may make the argument seem somewhat trivial, it nevertheless captures its essence. We consider
Bell's scenario with two local von Neumann measurements on two entangled qubits. Rather than
using the descriptors, we describe each qubit by the operational parameters defining the 
measurements which are performed on them. 

Suppose that two qubits are prepared in the singlet state 
$$
|\Psi\rangle=\frac{1}{\sqrt2}\left( |-1,1\rangle-|1, -1\rangle \right).
$$
The qubits are sent to two spatially separated parties, Alice and Bob. 
Each party independently chooses to perform a projective measurement, 
represented by the Bloch vectors $\vec a$ and $\vec b$, respectively.
Alice gets $s_A=\pm1$ and Bob $s_B=\pm1$. 
The outcomes of each measurement occur with equal probability, that is,
\be
P(s_A|\vec a)=P(s_B|\vec b)=1/2.
\ee
In the DH framework, this means that the qubits locally split
into two branches with equal weights.  On each side, this process occurs 
without the need to know what is happening on the other side, as a 
consequence of the {\it no signaling} property of quantum theory.
At this point of the experiment, no weight is attached on the
pair of outcomes $(s_A,s_B)$.
After the measurements, the parties decide to compare their results.
To do that, they have to meet at
the same place or send some communication through a classical channel. 
After multiple running of the experiment over different entangled 
states,
they find out that the outcomes, $s_A\in\{-1,1\}$ and $s_B\in\{-1,1\}$, 
are distributed according to the probability distribution
\be\label{ebit_prob}
P(s_A,s_B|\vec a,\vec b)=\frac{1}{4}\left(1-s_A s_B \vec a\cdot\vec b\right),
\ee
which violate the Bell inequalities. While this violation
leads to a break of local realism in a single-world scenario, the DH
framework evades this conclusion as follows. When Alice meets with
Bob, the branch in which she observed $s_A$ further divides into
two branches labeled by the value $s_B$ observed by Bob. The
probability of being paired with Bob observing $s_B$ given her observation
$s_A$ is 
\be
P(s_B|s_A,\vec a,\vec b)\equiv P(s_A,s_B|\vec a,\vec b)/P(s_A|\vec a). 
\ee
Thus, at the
meeting point, four branches emerge such that each one is weighted with an
amplitude whose modulus square is exactly $P(s_A,s_B|\vec a,\vec b)$.
By construction, these weights are not in conflict with the previous weight 
assignments.
Rephrasing in Saunders’ terms, the (infinite) multitudes of microstates of each party 
are paired in such a way that the fraction of realities experiencing the outcomes
$s_A$ and $s_B$ is equal to $P(s_A,s_B|\vec a,\vec b)$. 
All these processes occur locally, which is guaranteed by the no signaling conditions
\be
\begin{array}{l}
\sum_{s_B} P(s_A,s_B|\vec a,\vec b)=P(s_A|\vec a)  \\
\sum_{s_A} P(s_A,s_B|\vec a,\vec b)=P(s_B|\vec b).
\end{array}
\ee

It is worth recalling again Peres’ dictum that ``unperformed measurements have no result''.
The meeting is like a new measurement in which the pair of outcomes $(s_A, s_B)$ is
made definite. As long as the comparison of the results does not occur,
the parties can only claim their results (in their own reality), but the pairing
of the two local realities remains indefinite. The idea of postponing the build-up 
of the correlations is illustrated in Ref.~\cite{brassard} for nonlocal PR boxes
and further discussed in the general quantum case in Ref.~\cite{charles}.
The model in Ref.~\cite{brassard} simplifies the pairing rule at the
meeting point as the emerging branches have no weight. In particular,
the rule is identical to the rule discussed in Sec.~\ref{sec_locality}:
each instance of one party is paired with one instance of the
other party.

Thus,
the existence of many local realities allows us
to delay the buildup of the correlations until the parties
meet or communicate each other. 
It is fundamental to remark that, in the DH theory, each party has to 
carry the information about the performed measurement at the meeting
point in order to get the right pairing. Namely, the probability 
$P(s_A,s_B|\vec a,\vec b)$ must be known, which cannot be 
evaluated without knowing the Bloch vectors $\vec a$ and $\vec b$.
Thus, each party has
to communicate two real parameters, that is, infinite information.
Employing the descriptors of Ref.~\cite{deutsch}, the number of
real parameters would be $4^2-4=12$ per party~\cite{charles2}. The 
total number of independent parameters is $15$, since the descriptors of 
Alice are not completely independent of the descriptors of Bob.
This infinite amount of information is a consequence of the $\psi$-ontic 
nature of the theory.

\section{A model  with finite information flow}
\label{our_model_sec}
Employing the framework of Sec.~\ref{sec_locality}, let us show that two 
unweighted instances and a finite 
information flow are enough for simulating the outcomes of all the projective
measurements $\vec a$ and $\vec b$ on a maximally entangled state.
The proof is a simple adaptation of the Toner-Bacon model~\cite{toner}
to the ontological model of Sec.~\ref{sec_locality} with $m=2$ instances
per party. The Toner-Bacon model provides a classical simulation 
of the outcomes by using one bit of communication between the parties
and some shared randomness. Let us first review this single-world
model.

The classical protocol is as follows. Alice and Bob share two random
and independent unit vectors, $\vec x_0$ and $\vec x_1$, uniformly 
distributed over a sphere. First, Alice chooses to simulate the 
measurement $\vec a$ and generates an outcome $s_A$ such that
\be\label{s_A_eq}
s_A=\text{sign}(\vec a\cdot\vec x_0).
\ee
Then, she generates a number 
\be\label{n_A_eq}
n_A=\text{sign}(\vec a\cdot\vec x_0)\text{sign}(\vec a\cdot\vec x_1),
\ee
which is sent to Bob. Finally, Bob chooses to simulate the measurements
$\vec b$ and generates the outcome
\be\label{s_B_eq}
s_B=-\text{sign}\left[\vec b\cdot(\vec x_0+ n_A \vec x_1)\right].
\ee
Averaging on $\vec x_0$ and $\vec x_1$, it turns out that the outcomes
are generated according to Eq.~(\ref{ebit_prob}). Let us reproduce
the proof of Ref.~\cite{toner} by using the Kochen-Specker model~\cite{kochen}
as shortcut in the last step. It is sufficient to show that
\be\label{corr_eq}
\langle s_A s_B\rangle=-\vec a\cdot \vec b.
\ee
We have
\be
\begin{array}{l}
\langle s_A s_B\rangle= \frac{1}{(4\pi)^2}\sum_{n_A=\pm1}\int d^2 x_0 d^2 x_1 
\text{sign}(\vec a\cdot\vec x_0)      \vspace{1mm} \\
\text{sign} \left[-\vec b\cdot(\vec x_0+n_A \vec x_1) \right]
\frac{1+n_A \text{sign}(\vec a\cdot\vec x_0)(\vec a\cdot\vec x_1)}{2}.
\end{array}
\ee
First, we note that the two terms of the sum over $n_A$ are identical
(replace $\vec x_1$ with $-\vec x_1$ in the second term). 
Using the distributive property and a suitable swapping 
$\vec x_0\leftrightarrow \vec x_1$ , we have
\be
\langle s_A s_B\rangle= 
\frac{1}{8\pi^2}\int d^2 x_0 d^2 x_1 
\text{sign}(\vec a\cdot\vec x_0)      
\text{sign} \left[-\vec b\cdot(\vec x_0+\vec x_1) \right].
\ee
Integrating over $\vec x_1$ in spherical coordinates with $\vec b$
as the pole, we have
\be
\langle s_A s_B\rangle= \frac{1}{2\pi}\int d^2 x_0 
(-\vec b\cdot\vec x_0)\text{sign}(\vec a\cdot\vec x_0),
\ee
which can be recast in the form
\be
\langle s_A s_B\rangle=\frac{1}{\pi}\int d^2 x_0 
(-\vec b\cdot\vec x_0)\theta(\vec b\cdot\vec x_0)
\theta(\vec a\cdot\vec x_0)-(\vec a\rightarrow -\vec a)
\ee
where $\theta(x)$ is the Heaviside function. The first term 
appears in the Kochen-Specker model~\cite{kochen} and is
$-(1+\vec a\cdot\vec b)/2$. Thus, Eq.~(\ref{corr_eq}) is proved.

Now, let us reinterpret this model within the framework of 
Sec.~\ref{sec_locality} with $m=2$ instances per party.
The resulting model is schematically represented in Fig.~\ref{fig_model}.
The variables $\eta_A$ and $\eta_B$ in Sec.~\ref{sec_locality}
contain the integers $n_A$ and $n_B$, respectively, and a copy 
of the random variables $\vec x_0$ and $\vec x_1$. Initially,
the two instances are indistinguishable.
When the measurements are performed, 
each party branches into two distinct alternatives, labeled 
$A_{\pm1}$ and $B_{\pm1}$ for Alice and Bob, respectively. The
two parties are not allowed to communicate until they reach the
meeting point, where each alternative of Alice is paired to
one alternative of Bob according to some rule that we will determine.
First, we note that the Toner-Bacon model generates the same correlation
if $s_A$ and $s_B$ are both chosen with opposite sign.
Let us assume that alternatives $A_1$ and $A_{-1}$ get the outcome $s_A$ in
Eq.~(\ref{s_A_eq}) and its opposite, respectively. They both set
$n_A$ according to Eq.~(\ref{n_A_eq}). All these processes involve only
local information that is available to Alice. Thus, the ontic state
$\lambda_A$ of Eq.~(\ref{local_ontic}) evolves locally according to
Eq.~(\ref{local_causal_eqs}).
A complication arises in
deciding which outcomes are associated with Bob's alternatives. Associating
$s_B$ in Eq.~(\ref{s_B_eq}) to alternative $B_1$ would require to know
the value $n_A$, an information that is not available to Bob. If we
set $n_A$ equal to $1$ and, finally, we pair alternative $A_w$ with
$B_w$ at the meeting point, we end up making a mistake in the case 
$n_A$ were $-1$ and 
$\text{sign}\left[\vec b\cdot\vec x_+\right] \ne
\text{sign}\left[\vec b\cdot\vec x_-\right]$, where 
$\vec x_{\pm}\equiv \vec x_0\pm \vec x_1$. Crucially,
the mistake can be corrected during the pairing by knowing just one 
classical bit of information per measurement,
namely $n_A$ for Alice's measurement and
\be\label{n_B_eq}
n_B\equiv\text{sign}(\vec b\cdot\vec x_+)\text{sign}(\vec b\cdot\vec x_-)
\ee
for Bob's measurement.
Thus, let us assume that alternative $B_1$ and $B_{-1}$ generate
the outcomes 
\be\label{s_B_eq2}
s_B=-\text{sign}\left[\vec b\cdot(\vec x_0+\vec x_1)\right]
\ee
and $-s_B$, respectively.
Furthermore, both Bob's alternatives generate $n_B$ according 
to Eq.~(\ref{n_B_eq}),
which will be used to define the pairing. 
Also the ontic state $\lambda_B$ of Eq.~(\ref{local_ontic}) evolves locally 
according to Eq.~(\ref{local_causal_eqs}). Thus, the overall model is local.
Finally, Bob and Alice
reach the meeting point to compare their results. Bob and Alice carry
information about their outcome and the numbers $n_A$ and $n_B$,
respectively. Alternatives $A_{\pm 1}$ are paired with $B_{\pm1}$, unless
$n_A=n_B=-1$, in which case the pairing is swapped.
This leads to the following pairing rule
\be
\begin{array}{c}
(n_A,n_B)\ne(-1,-1) \Rightarrow  A_{\pm 1}\leftrightarrow B_{\pm1} \\
(n_A,n_B)=(-1,-1) \Rightarrow  A_{\pm 1}\leftrightarrow B_{\mp 1}.
\end{array}
\ee
Let us stress that there are only two instances after the pairing. If
$(n_A,n_B)\ne(-1,-1)$, then Alice and Bob agree that the pair of outcomes
is $(s_A,s_B)$ in one instance and $(-s_A,-s_B)$ in the other instance.
Conversely, if $(n_A,n_B)=(-1,-1)$, the pairing is flipped and the instances
are $(s_A,-s_B)$ and $(-s_A,s_B)$. It is worth to note that only
$2$ of the $4$ possible combinations of outcomes are realized in each
run. If there were $4$ unweighted instances deterministically realizing all 
the combinations, the probability of observing one pair of outcomes would be 
$1/4$ in each run. 
To match quantum theory using a finite number of unweighted instances, the final 
pair of outcomes must be stochastic -- that is, they must vary from run to run.
From the perspective of one realization of Alice and Bob, the probability
of getting a pair of outcomes $(s',s'')\in\{-1,1\}^2$ is obtained by a simple count of unweighted
instances. First, we compute the fraction of instances getting $(s',s'')$.
Then, we average over $\vec x_0$ and $\vec x_1$. There is no amplitude 
weight of the branches.

\begin{figure}[!h]
\usetikzlibrary{calc}
\begin{tikzpicture}[every node/.style={font=\small}]
  % Positions
  \coordinate (C) at (0,0);               % Circle center
  \coordinate (L) at (-1,2);              % Left square center
  \coordinate (R) at (1,2);               % Right square center
  \coordinate (T) at (0,4.5);             % Top rectangle center

  % Circle radius
  \def\circleradius{0.65}

  % Draw the bottom circle (smaller)
  \draw[thick] (C) circle (\circleradius cm);
  \node at ($(C)+(-0.2,0)$) {$\vec{x}_0$,};
  \node at ($(C)+(0.2,0)$) {$\vec{x}_1$};

  % Draw left and right squares
 \draw[thick] ($(L)+(-0.5,-0.5)$) rectangle ($(L)+(0.5,0.5)$);
 \draw[thick] ($(R)+(-0.5,-0.5)$) rectangle ($(R)+(0.5,0.5)$);

  % Arrows from circle to bottom-middle of squares
  \draw[->, thick] ($(C)+(-0.4,0.5)$) -- ($(L)+(0,-0.5)$);
  \draw[->, thick] ($(C)+(0.4,0.5)$) -- ($(R)+(0,-0.5)$);

  % Incoming a and b arrows
  \draw[->, thick] ($(L)+(-1,0)$) -- ($(L)+(-0.5,0)$);
  \node[left] at ($(L)+(-1,0)$) {$\vec a$};

  \draw[->, thick] ($(R)+(1,0)$) -- ($(R)+(0.5,0)$);
  \node[right] at ($(R)+(1,0)$) {$\vec b$};

  % Top rectangle labeled LNB
  \draw[thick] ($(T)+(-1,0.5)$) rectangle ($(T)+(1,-0.5)$);
  \node at (T) {meeting};

  % Arrows from squares to top rectangle
  \draw[->, thick] ($(L)+(0,0.5)$) -- ($(T)+(-0.2,-0.5)$);
  \node[above left] at ($(L)+(0.5,1.2)$) {$\vec{s}_A, n_A$};

  \draw[->, thick] ($(R)+(0,0.5)$) -- ($(T)+(0.2,-0.5)$);
  \node[above right] at ($(R)+(-0.5,1.2)$) {$\vec{s}_B, n_B$};

  \node[above] at ($(T)+(-0.0,0.5)$) {
	
	$
	\text{Local PR box} \left\{
	\begin{array}{cc}
	\text{communicated bits}  & \text{paired instances} \vspace{1mm}\\
	(n_A,n_B)\ne (-1,-1) \Rightarrow & (\pm s_{A},\pm s_{B}) \vspace{1mm}\\
	(n_A,n_B)=(-1,-1) \Rightarrow & (\pm s_{A},\mp s_{B})
	\end{array}
	\right.
	$
	};
     \draw[->, thick] 
    ($(T)+(1,0)$) -- ++(1.8,0) 
    |- ($(T)+(2.8,0.4)$);
\end{tikzpicture}
\caption{Local model simulating correlations of local projective measurements on
two maximally entangled qubits. Alice and Bob receive each two
random vectors $\vec x_0$ and $\vec x_1$ from a common source. Given projective
measurements $\vec a$ and $\vec b$, they locally generate $\{n_A, \vec s_A\equiv(s_A,-s_A)\}$
and
$\{n_B, \vec s_B\equiv(s_B,-s_B)\}$, respectively, by using only locally
available information. The numbers $s_A$, $n_A$, $s_B$ and $n_B$ are generated according to 
Eqs.~(\ref{s_A_eq},~\ref{n_A_eq},~\ref{s_B_eq2},~\ref{n_B_eq}), respectively.
Finally, they meet at a common point and each instance of
one party is paired with one instance of the other party according to the rule
in the figure. This results in two instances of the overall outcomes.}
\label{fig_model}
\end{figure}

Summarizing, the model is as follows. \newline
\begin{model} Simulation of local von Neumann measurements
with two instances and one communicated bit per party
in Bell's scenario of two maximally entangled qubits (Fig.~\ref{fig_model}).
\begin{itemize}
\item Alice and Bob receive the random vectors $\vec x_0$ and $\vec x_1$. 
\item In instance $A_{\pm1}$, Alice generates outcome $\pm s_A$ according to Eq.~(\ref{s_A_eq}),
given the measurement setting $\vec a$. She also sets $n_A$ by using Eq.~(\ref{n_A_eq}).
\item In instance $B_{\pm1}$, Bob generates outcome $\pm s_B$ according to Eq.~(\ref{s_B_eq2}),
given the measurement setting $\vec b$. He sets $n_B$ by using Eq.~(\ref{n_B_eq}).
\item Finally, they reach the meeting point. Instance $A_{\pm 1}$ is paired with
instance $B_{\pm 1}$ if $(n_A,n_B)\ne(-1,-1)$, otherwise the pairing is swapped.
\end{itemize}
At each run of the simulation, two instances of the overall outcomes are
generated. The quantum probabilities are obtained by selecting one of the two instances
at random with equal probability (assumption of equiprobability of instances in 
Sec.~\ref{multiple_instances_sec}).
\end{model}

It is interesting to note that the pairing depends only on $n_A$ and $n_B$. There is
no need to carry the initial random vectors $\vec x_0$ and $\vec x_1$, which are
used only for generating $s_A$, $s_B$, $n_A$ and $n_B$.

This model is mathematically equivalent to the model in Ref.~\cite{gisin},
but they differ in the interpretation. The latter is a single-world
model in which the parties share a nonlocal PR box as a resource for
generating the outcome. In our model, we have a ``local'' PR box which
is used for pairing locally the alternatives at the meeting point.
The statistics of the single-world model of Ref.~\cite{gisin} is recovered
by randomly taking one of the two paired alternatives in the two-instances
model. It is surprising that a simple and seemingly trivial conceptual
step separates our model from that in Ref.~\cite{gisin}, yet it has gone 
unnoticed until now.

It is interesting to note that a local single-world model could be constructed 
in a slightly different scenario. Instead of generating the correct outcomes 
at the moment measurements are performed, the task shifts to generating them 
at the meeting point. Each party only needs to communicate two bits. However,
including the whole process in this local description would 
lead to the paradoxical conclusion that a party's state -- and even their 
memory of past observations -- can suddenly change. In a many-instances framework, 
this macroscopic ``rewind'' and memory reconstruction is avoided by allowing 
two distinct realities to evolve independently.

\section{Discussion}
\label{discussion_sec}
The nonlocality arising in the framework of single-world ontological theories can 
lead the proponents of Deutsch-Hayden many-worlds theory to claim the failure of 
a probabilistic framework in which $\psi$ represents a mere state of knowledge.
The main purpose of this paper is to show 
with a proof of principle that the coexistence of multiple realities is
sufficient 
for a local account of the correlations of two maximally entangled qubits. 
There is no known proof that both determinism and the ontology  of the 
quantum states are necessary conditions for a local theory. Furthermore,
the drop of these two assumptions can lead to a more ``economical'' theory,
in which the information flow is made as small as possible and the
quantum probabilities are obtained by a simple count of unweighted
instances.

Let us discuss in detail the two key differences between the model in 
Sec.~\ref{our_model_sec} and the Deutsch-Hayden many-worlds theory -- namely, 
the way the probabilities come out and the role played by the quantum state. 
\newline
{\it Probabilities} -- In the scenario
simulated by the model, if Alice and Bob perform the measurements 
$\vec a$ and $\vec b$, then they observe outcome $s_A$ and $s_B$
with probability $(1-s_A s_B \vec a\cdot\vec b)/4$. The
many-worlds theory says that there are $4$ branches, one for each
outcome, weighted by an amplitude whose modulus square is the
observed probability. In Saunders' terms,
there is a multitude of branches of microstates
such that the ratio between the number of branches with outcomes $s_A$
and $s_B$
and the number of overall branches is equal to the observed probability.
Since this probability is a real number, this multitude is actually
infinite.

In our model, the story is quite different. There is a finite number of
{\it unweighted} instances (namely, two instances). The observed probability 
of the final pair of outcomes $(s_A,s_B)$  is obtained 
by taking the fraction of instances with outcome $(s_A,s_B)$ 
(the fraction is $0$ or $1/2$ in the specific model) 
and averaging over different runs. Like in the general framework of
Sec.~\ref{many_real_sec}, the different instances
have equal probability of being experienced by
an observer. Note that some outcomes are not realized in a
single run of the experiment (in MW theory, all the
outcomes are realized in each run). Indeed, two instances cannot realize all
the $4$ possible combinations of outcomes.
The remarkable point of the model is not the reduction from $4$ to $2$ 
realized branches, but the equiprobabilities of the two instances.
This is possible because of the stochasticity of the model.
As previously said, the derivation of the branch weights of many-worlds 
theory from a branch counting {\`a la} Saunders~\cite{saunders} would 
require a division of each of the $4$ branches into an infinite multitude of equally
probable instances (see also Sec.~\ref{limit_sec}, in which it is shown that
determinism is recovered in the limit of infinite instances in our framework).

Recalling the general discussion of Sec.~\ref{many_real_sec},
let us transliterate this feature of the model to an experiment generating 
the superposition 
\be\label{super_eq}
\psi_+ |+\rangle+\psi_- |-\rangle
\ee
where $|\pm\rangle$ are two macroscopically distinct states, and 
$|\psi_-|^2=10^{-40} |\psi_+|^2$. In many-worlds theory, both the
alternatives are realized, however we expect to experience $|+\rangle$ with 
near-certainty.  In our framework, we have a finite number $m$ of instances
such that the ratio between the number of instances associated with state 
$|-\rangle$ and the number of all the instances is equal to our expectation
of experiencing $|-\rangle$ in a single run of the experiment. If the
total number of instances is small ($m\ll 10^{40}$), this ratio is zero in most of the
runs. That is, the branch $|-\rangle$ is almost never realized. The ratio
turns out to be equal to $|\psi_-|^2$ only on average over
a large number of runs (like in any single-world ontological theory). 
This point of view is
in the middle between a single-world ontological theory, such as de Broglie-Bohm theory,
and many-worlds theory. In the former, there is some ontic variable saying
that the system is {\it always} realized in one of the two superposed states. 
The latter says that both the states are {\it always} realized in parallel 
worlds. As a single-world theory is nonlocal, as proved by Bell, we
may need more than one world for a local description, but between $1$ 
(single-world ontological theories) and infinity (many-worlds theory plus 
Saunders' branch counting) there is any integer greater than $1$!
\newline
{\it Role of the quantum state} -- The second difference is more subtle. In a $\psi$-epistemic theory, the 
ontic state does not contain the full information about the quantum state.
Rather, this information is encoded in the statistics of the ontic
state. Let us consider the quantum state emerging from Alice's measurement
and how its information is encoded in Deutsch-Hayden theory and our model. 
In the former, the ongoing
state, which has to be carried to the meeting point, is the 
vector $\vec a$ and the outcome associated with each branch.
Their knowledge allows us to infer the outgoing quantum state
after the measurement in each branch, meaning that the quantum state
is part of the ontology. In our model, this inference is not 
possible in a single run of the experiment. After the measurement,
Alice generates a number $n_A$ and associates an outcome to each
branch. This information and the knowledge of the shared random 
variables $\vec x_0$ and $\vec x_1$ enable us only to infer that
the quantum state $\vec a$ in branch $A_1$ after the measurement
is any unit vector such that Eqs.~(\ref{s_A_eq},\ref{n_A_eq}) are
fulfilled. Rather, the full information about quantum state is statistically
encoded in the ontic variables. This is necessary for making the
information flow from the parties to the meeting point 
finite~\cite{montina_comm_epist0,montina_comm_epist}. It is worth to
recall again that the whole ontic state after the measurements is not
actually required at the meeting point, since $\vec x_0$ and $\vec x_1$
are not involved in the pairing rule. 

\section{Conclusions}

In conclusion, we have introduced a general probabilistic framework that 
encompasses both single-world and many-worlds theories as limiting cases. 
This hybrid approach, which combines the randomness of single-world 
ontological theories with the multiplicity of coexisting realities found 
in many-worlds theory, offers a way to address the challenges inherent in 
each individual previous framework.
\begin{itemize}[left=0pt]
\item
On one hand, the coexistence of diverse realities
allows us to evade Bell’s theorem, which applies to single-world ontological 
theories. More broadly, it may also circumvent the contextuality requirement. 
For instance, a manifestation of preparation contextuality is the proof that 
ontological theories inherently break time symmetry~\cite{leifer1}.
However, with appropriate adaptations, our model provides a fully 
time-symmetric description of a scenario in which a qubit, in a 
maximally mixed state, is measured at two different times~\cite{monti_erase}. 
The time asymmetry proved in Ref.~\cite{leifer1} is transferred to the 
measurement devices and the subsequent comparison of results, which 
are inherently time-asymmetric processes. Similarly, the branching 
circumvents the recently proved theorem that measurements erase 
information~\cite{monti_erase}, which is another manifestation of
contextuality. An open question is whether 
a many-instances framework can also elude the Pusey-Barrett-Rudolph 
(PBR) theorem~\cite{pbr}, which implies that quantum mechanics is in conflict
with $\psi$-epistemic theories under a single-world premise and an assumption
of preparation independence (a kind of locality condition).  Note that 
this bypass of the theorem has never been considered so far, since 
Deutsch-Hayden theory is local but also $\psi$-ontic. It is also
interesting to note that the hypothesis of preparation independence
may be questioned
in the framework of nonlocal single-world ontological theories, but
it becomes mandatory in our framework, which aims to provide a locally
causal structure.
One possible way to circumvent the theorem is to assume that
multiple preparations may coexist, which is more drastic than assuming
that multiple outcomes coexist.
\item
On the other hand, the randomness of the framework makes it possible
to recover Born's rule from a simple count over a finite set of instances. 
Furthermore, it can lead to a reduction of
the information flow as much as possible by stripping the quantum
state of its ``ontic'' rank. 
Finally, if the quantum state represents our knowledge about a system 
rather than an element of reality, then it is possible that not all 
alternatives in a superposition are necessarily realized (as
discussed in Sec.~\ref{multiple_instances_sec} and illustrated by
our model).
If, according to prior knowledge, there existed a branch with a very faint 
amplitude in which Scipio was defeated in the Third Punic War, this would 
not necessarily imply that there is a world where Carthage rules over Rome. 
That is, the multiple instances of the framework discussed in 
Sec.~\ref{multiple_instances_sec} may not realize all possible branches.
\end{itemize}

Here, we have presented a proof of principle of this hybridization
by showing that local measurements on two maximally entangled 
qubits can be locally simulated with only two unweighted instances,
finite communication and shared randomness. It is plausible
that a generalization to many qubits and any quantum state will
require a blow-up of the shared variables. This increase of 
local resources occurs also in the Deutsch-Hayden many-worlds 
theory~\cite{deutsch}, in which the descriptors are represented 
by matrices acting on the whole Hilbert space of all involved
particles. The main difference, however, is that the shared random 
variables in our approach would
not depend on the state of the system -- they would be generated
before the ``game'' starts and would provide a stochastic background
over which the system evolves. A similar
idea has been suggested in Ref.~\cite{montina_non_markov} within a
single-world scenario. Our purpose has been to introduce a general
framework with a simple proof of principle, rather than a comprehensive
theory. Whether this hybrid approach can be extended to a general local 
theory distinct from DH theory remains an open and, in our opinion,
fascinating question. A negative answer to this question would 
suggest that every locally causal theory of quantum mechanics is
essentially the DH theory.

A generalization of the presented model to higher dimensional
systems can be obtained from the approximate model in Ref.~\cite{montina_approx} 
which extends the Toner-Bacon protocol to arbitrary dimensions.
Although this generalized model
is not exact, its error is very small (about $1\%$) for few qubits per party.
Its adaptation to our many-instances framework would provide a very accurate
model
for a sufficiently low dimension. This adaptation
can suggest improvements for reducing the error or finding exact
protocols.

From a broader perspective, this framework provides a generalized setting 
for quantum communication complexity. Rather than just quantifying 
the communication cost of classically simulating quantum entanglement,
one can consider a scenario involving a third party -- a referee -- 
who receives multiple alternative outcomes and determines how to pair them. 
In this setting, the question concerns not only the minimal communication cost
from the two parties to the referee, but also the minimal number of parallel 
unweighted instances required for a local simulation 
of quantum correlations. The standard communication complexity problem of
quantifying the required communication cost for simulating entanglement
is a subclass of this generalized setting with the slight variation that
the generation of one of the outcomes is postponed and delegated to
the referee. The subclass corresponds to
the situation in which there is one instance and one of the two parties
-- say Bob --
communicates to the referee the full information about his choice. Let us
underline that, in
this case, Bob is not supposed to generate the outcome, but delegates the
task to the referee.
A further generalization could allow for multiple parallel 
choices of measurements performed by Alice and Bob. Focusing only on one 
side of the problem, one can consider $\psi$-ontic models with infinite
communication and find the minimal number of instances for an exact local
simulation.
It can be interesting, for example, to investigate some kind of `pilot' theory
with a finite number of instances in which each instance is locally piloted by 
Deutsch-Hayden descriptors (rather than the wave-function like in pilot-wave 
theories). The task would be to find the minimal number
of instances and the pairing rules that lead to an exact simulation of quantum
correlations for more general entangled states.

In the context of this generalized setting in quantum communication complexity, 
there is an interesting aspect that is worth mentioning and offers perspectives
for future research. 
On one hand, it is known that 
the communication cost of simulating quantum correlations goes to
zero as the entangled state approaches a product state~\cite{renner}. 
Thus, if we adopt the communication cost as a measure of `nonlocality', weakly 
entangled states are less `nonlocal' than strongly entangled states (here, 
`nonlocality' refers to single-world scenarios).
On the other hand, maximally entangled states can be simulated using a single 
`nonlocal’ PR-box~\cite{gisin}, whereas weakly entangled states require more than 
one PR-box~\cite{brunner}. Thus, adopting the number of
PR-boxes as a measure of nonlocality, we obtain the opposite conclusion that
weakly entangled states are more `nonlocal' than maximally entangled states.
In our generalized setting of communication complexity, the task is not just
to evaluate the minimal communication cost, but also the minimal
number of parallel realities (instances) for a local account of quantum
correlations. 
It is entirely possible that these two quantities behave differently as functions 
of the `degree’ of entanglement. In particular, 
the number of required instances is likely related to the second measure of 
`nonlocality' (with PR-boxes). Thus, a local model with multiple instances could
potentially `explain' why weakly entangled states are more robust against the
detection loophole in an experimental Bell test~\cite{loophole}. 
Indeed, if more instances are required for a correct simulation, a single-world 
protocol is less likely to succeed by discarding some simulation runs instead of 
correctly pairing the instances. Thus,
our framework is not just relevant in quantum foundations, but suggests
new technical problems which have their own appeal for classical simulations 
of quantum entanglement and quantum communication. Our framework  may provide 
new insights into some of the counter-intuitive behaviors of entanglement.

Another line of future research can be inspired by the
approach of 
{\it many interacting worlds} in Ref.~\cite{hall}, which has been built on
ground of the dBB theory. As in our framework,
this approach employs a finite number of worlds but is equivalent to 
quantum mechanics only in the limit of infinite worlds. Thus, it is 
experimentally testable. The approach is nonlocal, so that one can wonder
whether a modification making it ``local'' is possible. For example, in
the search of such a modified approach, one could start from the descriptors, 
rather than the wave function.

\section*{Acknowledgments}
We wish to thank Charles B\'edard for useful comments and suggestions.
We also thank Lev Vaidman for pointing out that some statements in the 
previous version are not accepted by the entire Everettian community. Finally,
we are grateful to Sophie Berthelette for her careful reading of the 
manuscript.
The authors acknowledge support from the Swiss National Science Foundation (SNSF), 
project No. 200020- 214808.

\end{document}